\documentclass[amsmath,amssymb,showpacs,twocolumn]{revtex4}
\usepackage{graphicx}
\usepackage{amscd,amsmath,amsthm,amsfonts,amssymb}
\newtheorem{theorem}{Theorem}

\def\sech{{\rm sech}}

\def\PT{$\mathcal{PT}$}
\def\[{\begin{equation}}
\def\]{\end{equation}}

\begin{document}
\title{A new nonlocal nonlinear Schr\"odinger equation and its soliton solutions}
\author{Jianke Yang}
\affiliation{Department of Mathematics and Statistics, University of Vermont, Burlington, VT 05401, USA}
\begin{abstract}
A new integrable nonlocal nonlinear Schr\"odinger (NLS) equation with clear physical motivations is proposed. This equation is obtained from a special reduction of the Manakov system, and it describes Manakov solutions whose two components are related by a parity symmetry. Since the Manakov system governs wave propagation in a wide variety of physical systems, this new nonlocal equation has clear physical meanings. Solitons and multi-solitons in this nonlocal equation are also investigated in the framework of Riemann-Hilbert formulations. Surprisingly, symmetry relations of discrete scattering data for this equation are found to be very complicated, where constraints between eigenvectors in the scattering data depend on the number and locations of the underlying discrete eigenvalues in a very complex manner. As a consequence, general $N$-solitons are difficult to obtain in the Riemann-Hilbert framework. However, one- and two-solitons are derived, and their dynamics investigated. It is found that two-solitons are generally not a nonlinear superposition of one-solitons, and they exhibit interesting dynamics such as meandering and sudden position shifts. As a generalization, other integrable and physically meaningful nonlocal equations are also proposed, which include NLS equations of reverse-time and reverse-space-time types as well as nonlocal Manakov equations of reverse-space, reverse-time and reverse-space-time types.
\end{abstract}


\maketitle

\section{Introduction}
Integrable systems have been studied for over fifty years \cite{Ablowitz1981,Zakharov1984,Faddeev1987,Ablowitz1991,Yang2010}. The most familiar integrable systems are local equations, i.e., the solution's evolution depends only on the local solution value and its local space and time derivatives. The Korteweg-de Vries equation and the nonlinear Schr\"odinger (NLS) equation are such examples.

In the past few years, nonlocal integrable equations started to attract a lot of attention. The first such equation, as proposed by Ablowitz and Musslimani \cite{AblowitzMussPRL2013} as a new reduction of the Ablowitz-Kaup-Newell-Segur (AKNS) hierarchy \cite{AKNS1974}, is the NLS equation of reverse-space type,
\[ \label{e:NLSRX}
iq_t(x,t)+q_{xx}(x,t)+2\sigma q^2(x,t)q^*(-x,t)=0,
\]
where $\sigma=\pm 1$ is the sign of nonlinearity, and the asterisk * represents complex conjugation. This equation is distinctive because solution states at distant locations $x$ and $-x$ are directly coupled, reminiscent of quantum entanglement between pairs of particles.

Following the introduction of this equation, its properties have been extensively investigated \cite{AblowitzMussPRL2013,Xu2015,AblowitzMussNonli2016,WYY2016,HXLM2016,Gerdjikov2017,Stalin2017wrong,Zhang1,Peckan2017arxiv,Caudrelier,
BYJYrogue,YangNsoliton,Fengdarksoliton}. In addition, other nonlocal integrable equations have been reported
\cite{AblowitzMussPRE2014,Yan,Khara2015,Fokas2016,GerdjikovNwave2016,Chow,JZN2017,Lou2,AblowitzMussSAPM,ZhoudNLS,ZhouDS,
Zhu2,HePPTDS,Zhu3,BYJY2017,nonlocalFordy2017,AblowitzSG2017,BYnonlocalDS,Zhang2}. A transformation between many nonlocal and local equations has been discovered as well \cite{BYJY2017}.

From a mathematical point of view, studies of these nonlocal equations is interesting because these equations often feature new types of solution behaviors, such as finite-time solution blowup \cite{AblowitzMussPRL2013,BYJYrogue}, the simultaneous existence of solitons and kinks \cite{Zhu2}, the simultaneous existence of bright and dark solitons \cite{AblowitzMussPRL2013,RXNLS_dark}, and distinctive multi-soliton patterns \cite{YangNsoliton}. However, the physical motivations of these existing nonlocal equations are rather weak. Indeed, none of these equations was derived for a concrete physical system [even though the nonlocal equation (\ref{e:NLSRX}) above was linked to an unconventional system of magnetics \cite{PTNLSmagnetics}, it is not clear whether such an unconventional magnetics system is physically realizable]. This lack of physical motivation dampens the interest in these nonlocal equations from the broader scientific community.

In this article, we propose a new integrable nonlocal NLS equation which has clear physical meanings. This equation is
\[ \label{RXNLS0}
iu_t(x,t)+u_{xx}(x,t)+2\sigma \left[|u(x,t)|^2+|u(-x,t)|^2\right]u(x,t)=0,
\]
where $\sigma=\pm 1$. Here, the nonlocality is also of reverse-space type, where solutions at locations $x$ and $-x$ are directly coupled, similar to Eq. (\ref{e:NLSRX}). The difference from Eq.~(\ref{e:NLSRX}) is that the nonlinear terms are different. Here, the nonlinearity-induced potential $2\sigma [|u(x,t)|^2+|u(-x,t)|^2]$ is real and symmetric in $x$, which contrasts the previous equation (\ref{e:NLSRX}), where the nonlinearity-induced potential $2\sigma q(x,t)q^*(-x,t)$ is generally complex and parity-time-symmetric \cite{Yang_review}.

The new equation (\ref{RXNLS0}) will be derived from a special reduction of the Manakov system \cite{Manakov}. It is well known that the Manakov system governs nonlinear wave propagation in a great variety of physical situations, such as the interaction of two incoherent light beams \cite{Kivsharbook}, the transmission of light in a randomly birefringent optical fiber \cite{ManakovPMD1,ManakovPMD2}, and the evolution of two-component Bose-Einstein condensates \cite{BEC}. Thus, this new nonlocal equation governs nonlinear wave propagation in such physical systems under a certain constraint of the initial conditions, where the two components of the Manakov system are related by a parity symmetry. This physical interpretation can help us understand the solution behaviors in this nonlocal equation.

For this new integrable nonlocal equation, we will further study its bright solitons and multi-solitons in the framework of Riemann-Hilbert formulation (which is a modern treatment of the inverse scattering transform) \cite{Zakharov1984,Faddeev1987,Yang2010}. In this Riemann-Hilbert framework, the key to deriving general soliton solutions is to determine symmetry relations of the discrete scattering data. For the previous nonlocal NLS equation (\ref{e:NLSRX}) and two others of reverse-time and reverse-space-time types \cite{AblowitzMussPRL2013,AblowitzMussSAPM}, it was found in \cite{YangNsoliton} that those symmetry relations were very simple, and thus the general $N$-solitons in those equations were very easy to write down. However, for this new nonlocal equation (\ref{RXNLS0}), we will show that its symmetry relations of discrete scattering data are very complicated, because the constraints between eigenvectors in the scattering data depend on the number and locations of the underlying discrete eigenvalues in a very intricate way. Even though we do succeed in deriving these symmetry relations for one- and two-solitons, derivation of such relations for the general $N$-solitons is apparently very difficult, at least in the Riemann-Hilbert and inverse scattering framework. We are not aware of other integrable equations whose symmetry relations of the scattering data are so complicated, which makes this new nonlocal equation mathematically interesting and challenging. From the derived one- and two-soliton solutions, we find that two-solitons are generally not a nonlinear superposition of one-solitons, and they exhibit interesting dynamical patterns such as meandering and sudden position shifts. As a generalization of these results, we also propose other new integrable and physically meaningful nonlocal equations, such as the NLS equations of reverse-time and reverse-space-time types as well as nonlocal Manakov equations of reverse-space, reverse-time and reverse-space-time types.

\section{A new integrable nonlocal NLS equation}
The Manakov system
\begin{eqnarray}
&& iu_t+u_{xx}+2\sigma (|u|^2+|v|^2)u=0,   \label{Manakovu} \\
&& iv_t+v_{xx}+2\sigma (|u|^2+|v|^2)v=0,   \label{Manakovv}
\end{eqnarray}
where $\sigma=\pm 1$, is a ubiquitous nonlinear wave system which governs a wide variety of physical phenomena ranging from the interaction of two incoherent light beams \cite{Kivsharbook}, the transmission of light in a randomly birefringent optical fiber \cite{ManakovPMD1,ManakovPMD2}, and the evolution of two-component Bose-Einstein condensates \cite{BEC}. This system was shown by Manakov to be integrable \cite{Manakov} (see also \cite{Yang2010,Ablowitz_Trubatch_book}).

Now, we impose the solution constraint
\[  \label{constraint1}
v(x,t)=u(-x, t).
\]
Under this constraint, it is easy to see that the two equations in the Manakov system are consistent, and this system reduces to a single but nonlocal equation for $u(x,t)$ as
\[ \label{RXNLS}
iu_t(x,t)+u_{xx}(x,t)+2\sigma \left[|u(x,t)|^2+|u(-x,t)|^2\right]u(x,t)=0,
\]
which is the new nonlocal NLS equation (\ref{RXNLS0}) in the previous section.

The above derivation of this new nonlocal equation also reveals the physical interpretation of its solutions. Specifically, this equation describes solutions of the Manakov system under special initial conditions where $v(x, 0)=u(-x, 0)$. In this case, the $u(x,t)$ solution is governed by the nonlocal equation (\ref{RXNLS}), while the $v(x,t)$ solution is given in terms of $u(x,t)$ as $v(x,t)=u(-x, t)$. We emphasize that even though the Manakov system has been extensively studied before \cite{Manakov,Yang2010,Ablowitz_Trubatch_book}, its solutions with special initial conditions $v(x, 0)=u(-x, 0)$ have not received much attention. We just showed that these special solutions are governed by a single nonlocal equation (\ref{RXNLS}), which opens the door for studies of these solutions in the framework of Eq. (\ref{RXNLS}).

This new nonlocal equation is also integrable. To get its Lax pair, we recall that the Lax pair of the Manakov system (\ref{Manakovu})-(\ref{Manakovv}) are
\begin{eqnarray}
&& Y_x=(-i\zeta J+Q)Y, \label{Lax1} \\
&& Y_t=\left[-2i\zeta^2J+2\zeta Q+iJ(Q_x-Q^2)\right]Y,  \label{Lax2}
\end{eqnarray}
where
\[
J=\left(\begin{array}{ccc} 1 & 0 & 0 \\ 0 & 1 & 0 \\ 0 & 0 & -1\end{array}\right), \quad
Q=\left(\begin{array}{ccc} 0 & 0 & u \\
0 & 0 & v \\
-\sigma u^* & -\sigma v^* & 0 \end{array}\right).  \nonumber
\]
The Lax pair for the nonlocal equation (\ref{RXNLS}) are simply the above ones with $v(x,t)$ replaced by $u(-x, t)$ in view of the reduction
(\ref{constraint1}).

\section{Solitons and multi-solitons in the new nonlocal equation}

Since the new nonlocal equation (\ref{RXNLS}) is integrable, it is natural to seek its general soliton and multi-soliton solutions. Recall that this nonlocal equation is a reduction of the Manakov system. Thus, its solitons are a part of Manakov solitons. But what Manakov solitons satisfy this nonlocal equation? This is actually a nontrivial question. The present situation is similar to the previous nonlocal NLS equation (\ref{e:NLSRX}). Even though that equation was a reduction of the well-known coupled $q$-$r$ system in the AKNS hierarchy \cite{Yang2010,AblowitzMussPRL2013,AKNS1974}, its solutions were still not obvious, which prompted a lot of studies on that equation in the past few years \cite{AblowitzMussPRL2013,Xu2015,AblowitzMussNonli2016,WYY2016,HXLM2016,Gerdjikov2017,Stalin2017wrong,Zhang1,
Peckan2017arxiv,Caudrelier,BYJYrogue,YangNsoliton,Fengdarksoliton}.
In this article, we only consider bright-soliton solutions, which exist under focusing nonlinearity; thus we set $\sigma=1$ below.

In a previous article \cite{YangNsoliton}, we derived general $N$-solitons in the previous nonlocal NLS equation (\ref{e:NLSRX}) and two others of reverse-time and reverse-space-time types, which were reduced from the $q$-$r$ system in the AKNS hierarchy \cite{AblowitzMussPRL2013,AKNS1974,AblowitzMussSAPM}. That derivation was set in the Riemann-Hilbert framework. Starting from the general $N$-soliton solutions of the $q$-$r$ system and deriving symmetry relations of the discrete scattering data for those nonlocal equations, general $N$-solitons were then obtained. In that approach, derivation of symmetry relations of the scattering data was the key. It turns out that those symmetry relations were simple (as in all common integrable systems we are aware of). Thus, general $N$-solitons in those nonlocal equations were easy to write down.

In this article, we follow a similar approach. Since the new nonlocal equation (\ref{RXNLS}) is a reduction from the Manakov system, we will start from the general soliton solutions of the Manakov system in the Riemann-Hilbert formulation. As before, the key to obtaining solitons in this new nonlocal equation is to derive symmetry relations of its discrete scattering data, which we will do below. It turns out that these symmetry relations are surprisingly complicated for this new nonlocal equation, which makes derivations of its general $N$-solitons more difficult.

General $N$-solitons of the Manakov system (\ref{Manakovu})-(\ref{Manakovv}) are well known \cite{Yang2010,Manakov,Ablowitz_Trubatch_book}. From the Riemann-Hilbert approach, such solitons can be expressed as \cite{Yang2010}
\begin{equation}  \label{uvmanakov}
\left(\begin{array}{c} u(x, t)  \\ v(x, t) \end{array}\right)
=2i\sum\limits_{j,k=1}^N \left(\begin{array}{c} \alpha_j  \\
\beta_j \end{array}\right) e^{\theta_j-\theta_k^* }\left( {M^{-1}
}\right)_{jk},
\end{equation}
where $M$ is a $N\times N$ matrix whose elements are given by
\begin{equation*}
M_{jk}=\frac{1}{{\zeta_j^*-\zeta_k }}\left[e^{-\left(
\theta_j^*+\theta_k \right)}+(\alpha_j^*\alpha_k+\beta_j^*\beta_k)
e^{\theta_j^*+ \theta_k} \right],
\end{equation*}
\begin{equation} \label{e:thetak}
\theta_k=-i\zeta_k x-2i\zeta_k^2 t,
\end{equation}
$\zeta_k$ are complex numbers in the upper half plane $\mathbb{C}_+$, and $\alpha_k, \beta_k$ are arbitrary complex constants. In the language of inverse scattering, the discrete scattering data of these solitons is $\{\zeta_k, \textbf{w}_{k0}, 1\le k \le N\}$, where $\zeta_k$ are  zeros of the underlying Riemann-Hilbert problem (which are assumed to be simple), and
\begin{equation*}
\textbf{w}_{k0}=(\alpha_k, \beta_k, 1)^T
\end{equation*}
is the associated eigenvector at the Riemann-Hilbert zero $\zeta_k$. Here, the superscript `$T$' represents transpose of a vector, and the eigenvector $\textbf{w}_{k0}$ has been scaled so that its last element is unity.

The nonlocal equation (\ref{RXNLS}) is obtained from the Manakov equations under the solution reduction (\ref{constraint1}). This solution reduction becomes a potential constraint in the scattering problem (\ref{Lax1}), which induce symmetry conditions on the scattering data. Imposing these symmetry conditions of the scattering data in the above Manakov solitons, the resulting solutions would be solitons of the nonlocal equation (\ref{RXNLS}).

Note that the Manakov $N$-solitons (\ref{uvmanakov}) above already incorporated symmetry conditions of the scattering data between eigenvalues in the upper and lower complex planes $\mathbb{C}_+$ and $\mathbb{C}_-$, which appear as complex-conjugate pairs \cite{Zakharov1984,Yang2010}. Such symmetry conditions are valid for all Manakov solitons. For the present nonlocal equation (\ref{RXNLS}), we only need to determine symmetry conditions of scattering data for eigenvalues in the upper complex plane $\mathbb{C}_+$, which are induced by the new potential reduction (\ref{constraint1}). These symmetry conditions are presented in the following theorem.

\begin{theorem} \label{Theorem1}
For the nonlocal NLS equation (\ref{RXNLS}), if $\zeta\in \mathbb{C}_+$ is a discrete eigenvalue, so is $\hat{\zeta}\equiv -\zeta^* \in \mathbb{C}_+$. Thus, eigenvalues in the upper complex plane are either purely imaginary, or appear as $(\zeta, -\zeta^*)$ pairs.
Symmetry relations on their eigenvectors depend on the number and locations of these eigenvalues. For the one- and two-solitons (with a single and double eigenvalues in $\mathbb{C}_+$ respectively), these symmetry relations are given below.
\begin{enumerate}
\item For a single purely imaginary eigenvalue $\zeta_1=i\eta$, with $\eta>0$, its eigenvector is of the form
\[ \label{wsym2a}
\textbf{w}_{10}=\left[2^{-1/2}e^{i\gamma}, \; 2^{-1/2}e^{i\gamma}, \; 1\right]^T,
\]
where $\gamma$ is an arbitrary real constant.
\item For two purely-imaginary eigenvalues $\zeta_1=i\eta_1$ and $\zeta_2=i\eta_2$, with $\eta_1, \eta_2 >0$,
their eigenvectors $\textbf{w}_{10}=(\alpha_1, \beta_1, 1)^T$ and $\textbf{w}_{20}=(\alpha_2, \beta_2, 1)^T$ are related as
\begin{eqnarray}
&& |\alpha_1|^2+|\beta_1|^2=|\alpha_2|^2+|\beta_2|^2,  \label{Th2a}\\
&& g^2\left[(|\alpha_1|^2+|\beta_1|^2)^2-1\right]  \nonumber \\
&& \hspace{0.8cm} =(1-g^2)\left(1-|\alpha_1^*\alpha_2+\beta_1^*\beta_2|^2\right), \label{Th2b} \\
&& \beta_1=(1+g)(\alpha_1\alpha_2^*+\beta_1\beta_2^*)\alpha_2  \nonumber \\
&& \hspace{0.8cm}  -g(|\alpha_2|^2+|\beta_2|^2)\alpha_1,  \label{Th2c} \\
&& \beta_2=g(|\alpha_1|^2+|\beta_1|^2)\alpha_2   \nonumber \\
&& \hspace{0.8cm}  +(1-g)(\alpha_1^*\alpha_2+\beta_1^*\beta_2)\alpha_1,  \label{Th2d}
\end{eqnarray}
where
\[ \label{def:g}
g\equiv (\eta_1+\eta_2)/(\eta_2-\eta_1).
\]
These equations admit solutions for $\textbf{w}_{10}$ and $\textbf{w}_{20}$ if and only if
$|\alpha_1^*\alpha_2+\beta_1^*\beta_2|\le 1$, and the admitted solutions have four free real parameters (not counting the eigenvalue parameters $\eta_1$ and $\eta_2$).

\item For two non-purely-imaginary eigenvalues $(\zeta_1, \zeta_2) \in \mathbb{C}_+$,
where $\zeta_2=-\zeta^*_1$, their eigenvectors $\textbf{w}_{10}=(\alpha_1, \beta_1, 1)^T$ and $\textbf{w}_{20}=(\alpha_2, \beta_2, 1)^T$ are related as
\[ \label{wsym2c}
\left(\begin{array}{c} \alpha_2 \\ \beta_2 \end{array}\right)=S \left(\begin{array}{c} \alpha_1 \\ \beta_1 \end{array}\right),
\]
where
\begin{eqnarray*}
&& S=\left(\frac{1}{\zeta_1^*-\zeta_1}-\frac{1}{2\zeta_1^*}\right) \times  \\
&& \hspace{0.5cm} 
\left(\begin{array}{cc}
-\frac{\alpha_1^*\beta_1}{2\zeta_1^*} & \frac{|\alpha_1|^2+|\beta_1|^2}{\zeta_1^*-\zeta_1}-\frac{|\beta_1|^2}{2\zeta_1^*} \\
\frac{|\alpha_1|^2+|\beta_1|^2}{\zeta_1^*-\zeta_1}-\frac{|\alpha_1|^2}{2\zeta_1^*} & -\frac{\alpha_1\beta_1^*}{2\zeta_1^*}
\end{array}\right)^{-1},
\end{eqnarray*}
and $\alpha_1, \beta_1$ are free complex constants.
\end{enumerate}
\end{theorem}

\noindent
\textbf{Proof.} The nonlocal NLS equation (\ref{RXNLS}) is reduced from the Manakov system under the solution reduction $v(x,t)=u(-x,t)$. In this case, the initial potential matrix
\begin{equation*}
Q(x,0)=\left(\begin{array}{ccc} 0 & 0 & u(x,0) \\
0 & 0 & u(-x,0) \\
-u^*(x,0) & -u^*(-x,0) & 0 \end{array}\right)
\end{equation*}
admits the following two symmetries
\begin{equation*}
Q^\dagger(x,0)=-Q(x,0), \hspace{0.2cm}  Q(-x,0)=-P^{-1}Q(x,0)P,
\end{equation*}
where
\[
P=\left(\begin{array}{ccc} 0 & 1 & 0 \\ 1 & 0 & 0 \\ 0 & 0 & -1\end{array}\right),   \nonumber
\]
and the dagger $\dagger$ represents Hermitian (i.e., conjugate transpose). The first potential symmetry $Q^\dagger=-Q$ is valid for the general Manakov system, and we only need to consider the second potential symmetry and its consequences.

Switching $x \to -x$ in the scattering equation (\ref{Lax1}) and utilizing the second potential symmetry, we get
\[ \label{e:Y2sym}
\left[PY(-x)\right]_x=\left[\zeta J+Q(x)\right]\left[PY(-x)\right].
\]
This means that, if $\zeta$ is a discrete eigenvalue of the scattering problem (\ref{Lax1}), so is $-\zeta$. But it is known for the general Manakov system that eigenvalues to the scattering problem (\ref{Lax1}) come in conjugate pairs $(\zeta, \zeta^*)$. Thus, if $-\zeta$ is an eigenvalue, so is $-\zeta^*$. This proves the eigenvalue symmetry in Theorem \ref{Theorem1}.

It is important to notice that, although we can show $-\zeta^*$ would be an eigenvalue so long as $\zeta$ is, there is no simple relation between their eigenfunctions, and thus one cannot obtain a simple symmetry relation between their eigenvectors in the scattering data. Eigenfunctions for $\zeta$ and $-\zeta$ are directly related in view of Eq. (\ref{e:Y2sym}). But $-\zeta$ is in the opposite half plane of $\zeta$, and the adjoint eigenfunction at $-\zeta$, which we need \cite{Zakharov1984,Yang2010}, is not available.

To prove symmetry relations of eigenvectors for one- and two-solitons in Theorem \ref{Theorem1}, we utilize the connection between these eigenvectors and Riemann-Hilbert-based $N$-soliton solutions (\ref{uvmanakov}) of the Manakov system. By imposing the condition $v(x,t)=u(-x,t)$ on the Manakov solitons, we will be able to derive symmetry conditions of eigenvectors for the nonlocal NLS equation (\ref{RXNLS}).

First, we consider one-solitons, where there is a single purely imaginary eigenvalue $\zeta_1=i\eta \in \mathbb{C}_+$, with $\eta>0$. In this case, the one-Manakov-soliton from Eq. (\ref{uvmanakov}) can be rewritten as
\begin{equation*}
\left(\begin{array}{c} u(x, t)  \\ v(x, t) \end{array}\right)
=\left(\begin{array}{c} \alpha_1  \\
\beta_1 \end{array}\right) \frac{4\eta e^{4i\eta^2 t}}{e^{-2\eta x}+(|\alpha_1|^2+|\beta_1|^2)e^{2\eta x}}.
\end{equation*}
By requiring $v(x,t)=u(-x,t)$, we get the conditions
\[
\beta_1=\alpha_1 (|\alpha_1|^2+|\beta_1|^2), \quad \alpha_1=\beta_1 (|\alpha_1|^2+|\beta_1|^2).  \nonumber
\]
Hence,
\[
|\alpha_1|^2+|\beta_1|^2=1, \quad \alpha_1=\beta_1,    \nonumber
\]
and $|\alpha_1|^2=1/2$. Writing $\alpha_1=2^{-1/2}e^{i\gamma}$, where $\gamma$ is a real constant, the resulting eigenvector $\textbf{w}_{10}$ is then as given in Eq. (\ref{wsym2a}).

Next, we consider two-solitons, where there are two complex eigenvalues $\zeta_1, \zeta_2 \in \mathbb{C}_+$. In this case, the general two-Manakov-solitons from Eq. (\ref{uvmanakov}) can be rewritten as
\begin{eqnarray}
&& \hspace{-0.2cm} u(x,t)=\frac{2i}{\det(M)}\left[A_1 e^{\theta_1-\theta_1^*-(\theta_2+\theta_2^*)} +A_2 e^{\theta_1-\theta_1^*+\theta_2+\theta_2^*} \right. \nonumber \\
&& \hspace{1.1cm}\left. +A_3 e^{\theta_1+\theta_1^*+\theta_2-\theta_2^*}+A_4 e^{-(\theta_1+\theta_1^*)+\theta_2-\theta_2^*}\right],  \label{e:2Manakov1}
\end{eqnarray}
\begin{eqnarray}
&& \hspace{-0.2cm} v(x,t)=\frac{2i}{\det(M)}\left[B_1 e^{\theta_1-\theta_1^*-(\theta_2+\theta_2^*)} +B_2 e^{\theta_1-\theta_1^*+\theta_2+\theta_2^*} \right. \nonumber \\
&& \hspace{1.1cm}\left. +B_3 e^{\theta_1+\theta_1^*+\theta_2-\theta_2^*}+B_4 e^{-(\theta_1+\theta_1^*)+\theta_2-\theta_2^*}\right],
\label{e:2Manakov2}
\end{eqnarray}
where
\begin{eqnarray*}
&& \det(M)=C_1 e^{-(\theta_1+\theta_1^*+\theta_2+\theta_2^*)} + C_2 e^{\theta_1+\theta_1^*+\theta_2+\theta_2^*}   \\
&& \hspace{1.1cm} + C_3 e^{\theta_1+\theta_1^*-(\theta_2+\theta_2^*)} + C_4 e^{-(\theta_1+\theta_1^*)+\theta_2+\theta_2^*}  \\
&& \hspace{1.1cm} + C_5 e^{\theta_1-\theta_1^*-(\theta_2-\theta_2^*)} + C_5^* e^{-(\theta_1-\theta_1^*)+\theta_2-\theta_2^*}, 
\end{eqnarray*}
$\theta_k$ is given in Eq. (\ref{e:thetak}), and coefficients $A_k, B_k, C_k$ are certain functions of $\zeta_1, \zeta_2, \alpha_1, \alpha_2, \beta_1, \beta_2$ whose expressions are given below:
\[
A_1=\left(\frac{1}{\zeta_2^*-\zeta_2}-\frac{1}{\zeta_1^*-\zeta_2}\right)\alpha_1,  \nonumber
\]
\[
A_2=\frac{\alpha_1 (|\alpha_2|^2+|\beta_2|^2)}{\zeta_2^*-\zeta_2}-\frac{\alpha_2 (\alpha_1\alpha_2^*+\beta_1\beta_2^*)}{\zeta_2^*-\zeta_1},
\nonumber
\]
\[
A_3=\frac{\alpha_2 (|\alpha_1|^2+|\beta_1|^2)}{\zeta_1^*-\zeta_1}-\frac{\alpha_1 (\alpha_1^*\alpha_2+\beta_1^*\beta_2)}{\zeta_1^*-\zeta_2},
\nonumber
\]
\[
A_4=\left(\frac{1}{\zeta_1^*-\zeta_1}-\frac{1}{\zeta_2^*-\zeta_1}\right)\alpha_2,
\nonumber
\]
\[
B_1=\left(\frac{1}{\zeta_2^*-\zeta_2}-\frac{1}{\zeta_1^*-\zeta_2}\right)\beta_1,
\nonumber
\]
\[
B_2=\frac{\beta_1 (|\alpha_2|^2+|\beta_2|^2)}{\zeta_2^*-\zeta_2}-\frac{\beta_2 (\alpha_1\alpha_2^*+\beta_1\beta_2^*)}{\zeta_2^*-\zeta_1},
\nonumber
\]
\[
B_3=\frac{\beta_2 (|\alpha_1|^2+|\beta_1|^2)}{\zeta_1^*-\zeta_1}-\frac{\beta_1 (\alpha_1^*\alpha_2+\beta_1^*\beta_2)}{\zeta_1^*-\zeta_2},
\nonumber
\]
\[
B_4=\left(\frac{1}{\zeta_1^*-\zeta_1}-\frac{1}{\zeta_2^*-\zeta_1}\right)\beta_2,
\nonumber
\]
\[
C_1=\frac{1}{(\zeta_1^*-\zeta_1)(\zeta_2^*-\zeta_2)}+\frac{1}{|\zeta_1^*-\zeta_2|^2},
\nonumber
\]
\begin{eqnarray*}
C_2=\frac{\left(|\alpha_1|^2+|\beta_1|^2\right)\left(|\alpha_2|^2+|\beta_2|^2\right)}{(\zeta_1^*-\zeta_1)(\zeta_2^*-\zeta_2)}+ \frac{|\alpha_1^*\alpha_2+\beta_1^*\beta_2|^2}{|\zeta_1^*-\zeta_2|^2},
\end{eqnarray*}
\[
C_3=\frac{|\alpha_1|^2+|\beta_1|^2}{(\zeta_1^*-\zeta_1)(\zeta_2^*-\zeta_2)},
\nonumber
\]
\[
C_4=\frac{|\alpha_2|^2+|\beta_2|^2}{(\zeta_1^*-\zeta_1)(\zeta_2^*-\zeta_2)},
\nonumber
\]
\[
C_5=\frac{\alpha_1\alpha_2^*+\beta_1\beta_2^*}{|\zeta_1^*-\zeta_2|^2}.
\nonumber
\]
For two-solitons, there are two cases to consider.

\noindent
(1) If the two eigenvalues $\zeta_1$ and $\zeta_2$ are purely imaginary, i.e.,
\[ 
\zeta_1=i\eta_1, \quad \zeta_2=i\eta_2,    \nonumber
\]
with $\eta_1, \eta_2 >0$, then
\[
\theta_k=\eta_k x +2i\eta_k^2t, \quad \theta_k+\theta_k^*=2\eta_k x, \quad \theta_k-\theta_k^*=4i\eta_k^2t.   \nonumber
\]
In this case, when $x \to -x$,
\[
\theta_k+\theta_k^* \to -(\theta_k+\theta_k^*), \quad \theta_k-\theta_k^* \to \theta_k-\theta_k^*.   \nonumber
\]
Thus, by cross multiplication of the ratio expressions for $u(-x,t)$ and $v(x,t)$ from (\ref{e:2Manakov1})-(\ref{e:2Manakov2})
and requiring exponentials of the same power to match, we find that the necessary and sufficient conditions for $v(x,t)=u(-x, t)$ are
\[ \label{e:ABcond1}
A_1=B_2, \quad A_2=B_1, \quad A_3=B_4, \quad A_4=B_3,
\]
\[
C_1=C_2, \quad C_3=C_4.
\]
The requirement of $C_3=C_4$ directly leads to Eq. (\ref{Th2a}) in Theorem \ref{Theorem1}, and the requirement of $C_1=C_2$ leads to Eq. (\ref{Th2b}). Under these two requirements on $C_k$'s, we find that only two of the four conditions for $A_k$'s and $B_k$'s in Eq. (\ref{e:ABcond1}) are independent, i.e., if two of them are satisfied, then the other two would be satisfied automatically. When we choose the two conditions as $A_2=B_1$ and $A_3=B_4$, these conditions would lead to equations (\ref{Th2c})-(\ref{Th2d}).

Later in Sec. \ref{sec_imageig}, we will explicitly solve the four equations (\ref{Th2a})-(\ref{Th2d}), and show that they admit solutions for $\textbf{w}_{10}$ and $\textbf{w}_{20}$ if and only if $|\alpha_1^*\alpha_2+\beta_1^*\beta_2|\le 1$. In addition, the admitted solutions have four free real parameters (not counting the eigenvalue parameters $\eta_1$ and $\eta_2$).

\vspace{0.2cm}
\noindent
(2) If the two eigenvalues $\zeta_1$ and $\zeta_2$ are not purely imaginary, then $\zeta_2=-\zeta^*_1$. In this case,
\[
\theta_1=-i\zeta_1 x -2i\zeta_1^2t, \quad \theta_2=i\zeta_1^* x -2i\zeta_1^{*2}t;   \nonumber
\]
thus, 
\[
\theta_1+\theta_2^*=-2i\zeta_1x, \quad  \theta_1-\theta_2^*=-4i\zeta_1^2t.    \nonumber
\]
Then, as $x \to -x$,
\[
\theta_1+\theta_2^* \to -(\theta_1+\theta_2^*), \quad \theta_1-\theta_2^* \to \theta_1-\theta_2^*.   \nonumber
\]
Recalling the expressions of $u(x,t)$ and $v(x,t)$ in Eqs. (\ref{e:2Manakov1})-(\ref{e:2Manakov2}), we find that in order for $v(x,t)=u(-x, t)$, the necessary and sufficient conditions now are
\[ \label{Th2case3AB}
A_1=B_3, \quad A_2=B_4, \quad A_3=B_1, \quad A_4=B_2,
\]
and
\[ \label{Th2case3C}
C_1=C_2, \quad C_5=C_5^*.
\]
The $A_1=B_3$ and $A_3=B_1$ conditions are
\begin{eqnarray*}
&& \frac{\beta_2(|\alpha_1|^2+|\beta_1|^2)}{\zeta_1^*-\zeta_1}-\frac{\beta_1(\alpha_1^*\alpha_2+\beta_1^*\beta_2)}{2\zeta_1^*}   \\  
&& = \left(\frac{1}{\zeta_1^*-\zeta_1}-\frac{1}{2\zeta_1^*}\right)\alpha_1
\end{eqnarray*}
and
\begin{eqnarray*}
&& \frac{\alpha_2(|\alpha_1|^2+|\beta_1|^2)}{\zeta_1^*-\zeta_1}-\frac{\alpha_1(\alpha_1^*\alpha_2+\beta_1^*\beta_2)}{2\zeta_1^*}  \\
&& =\left(\frac{1}{\zeta_1^*-\zeta_1}-\frac{1}{2\zeta_1^*}\right)\beta_1,     
\end{eqnarray*}
which can be rewritten as equations (\ref{wsym2c}) in Theorem \ref{Theorem1}. Remarkably, we find that when $(\alpha_2, \beta_2)$ are related to $(\alpha_1, \beta_1)$ by Eq. (\ref{wsym2c}), all the other conditions in (\ref{Th2case3AB})-(\ref{Th2case3C}) are automatically satisfied. This completes the proof of Theorem \ref{Theorem1}. $\Box$

\textbf{Remark 1.} Theorem \ref{Theorem1} shows that for the nonlocal NLS equation (\ref{RXNLS}), symmetry relations of eigenvectors in the scattering data are very complicated, because such relations depend on the number and locations of eigenvalues in a highly nontrivial way. Given the complexity of these symmetry relations for two-solitons, such relations for three and higher solitons are expected to be even more complicated. This poses a challenge for deriving general $N$-solitons in Eq. (\ref{RXNLS}), at least in the Riemann-Hilbert framework.

\section{Soliton dynamics in the nonlocal NLS equation}

In this section, we examine dynamics of one- and two-solitons of Eq. (\ref{RXNLS}) as presented in Theorem \ref{Theorem1}.

\subsection{Single solitons} \label{onesoliton}
Single solitons in the nonlocal NLS equation (\ref{RXNLS}) can be obtained from the single Manakov-soliton (\ref{uvmanakov}) with one purely imaginary eigenvalue $\zeta_1=i\eta$ $(\eta>0)$ and with its eigenvector $\textbf{w}_{10}$ given by Eq. (\ref{wsym2a}) in Theorem \ref{Theorem1}. This soliton is
\[
u(x,t)=\sqrt{2}\hspace{0.03cm} \eta \hspace{0.03cm} e^{4i\eta^2t+i\gamma}\sech\left(2\eta x\right),   
\]
where $\gamma$ is a free real parameter. Since the nonlocal NLS equation (\ref{RXNLS}) is phase-invariant, the above soliton is equivalent to
\[ \label{1solitonRX}
u(x,t)=\sqrt{2}\hspace{0.03cm} \eta \hspace{0.03cm} e^{4i\eta^2t}\sech \left(2\eta x\right),
\]
which is shown in Fig. 1. This soliton is stationary with constant amplitude, and is symmetric in $x$.

\begin{figure}[htbp]
\begin{center}
\includegraphics[width=0.48\textwidth]{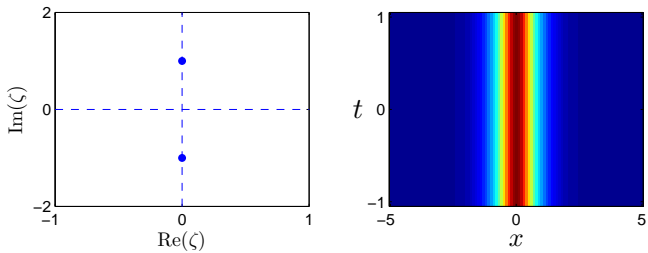}
\caption{The single soliton (\ref{1solitonRX}) in the nonlocal NLS equation (\ref{RXNLS}) with $\eta=1$. Left panel: positions of eigenvalues. Right panel: graph of solution $|u(x,t)|$. } \label{f:fig1}
\end{center}
\end{figure}

\subsection{Two-solitons with purely imaginary eigenvalues} \label{sec_imageig}
These solitons are obtained from the two-Manakov-solitons (\ref{uvmanakov}) with two purely imaginary eigenvalues in $\mathbb{C}_+$, and with eigenvectors $\textbf{w}_{10}$, $\textbf{w}_{20}$ satisfying the equations (\ref{Th2a})-(\ref{Th2d}) in Theorem \ref{Theorem1}. Below, we solve these four equations explicitly.

First, we introduce the notations
\[ 
p\equiv |\alpha_1|^2+|\beta_1|^2, \quad q\equiv \alpha_1^*\alpha_2+\beta_1^*\beta_2,    \nonumber
\]
and
\[
q\equiv r_0e^{i\gamma_0}, \quad \alpha_1\equiv r_1e^{i\gamma_1}, \quad \alpha_2\equiv r_2e^{i\gamma_2},   \nonumber
\]
where $r_0, r_1, r_2 \: (\ge 0)$ are amplitudes of complex numbers $q, \alpha_1, \alpha_2$, and $\gamma_0, \gamma_1, \gamma_2$ their phases.

Before solving equations (\ref{Th2a})-(\ref{Th2d}), we notice that they admit two invariances, i.e., if
\[
\alpha_1 \to \alpha_1 e^{i\widehat{\gamma}_1}, \hspace{0.1cm} \beta_1\to \beta_1 e^{i\widehat{\gamma}_1}, \hspace{0.1cm}
\alpha_2 \to \alpha_2 e^{i\widehat{\gamma}_2}, \hspace{0.1cm} \beta_2\to \beta_2 e^{i\widehat{\gamma}_2},    \nonumber
\]
where $\widehat{\gamma}_1$ and $\widehat{\gamma}_2$ are arbitrary real constants, then these equations remain invariant.
Thus, the phases $\gamma_1, \gamma_2$ of parameters $\alpha_1$ and $\alpha_2$ are free real constants.

To solve equations (\ref{Th2a})-(\ref{Th2d}), it is convenient to parameterize their solutions in terms of $q$, i.e., $r_0$ and $\gamma_0$, which are two additional free real constants. We will show that solutions exist if and only if $|q| \le 1$.

For given $q$, we can get $p$ from Eq. (\ref{Th2b}) as
\[
p=\sqrt{\frac{1+(g^2-1)|q|^2}{g^2}}.   \nonumber
\]
Recall from the definition of $g$ in Eq. (\ref{def:g}) that $g$ is real and $|g|>1$. Thus, the quantity under the square root in the above expression is always positive. After the $p$ and $q$ values are available, we see from Eqs. (\ref{Th2a}) and (\ref{Th2c})-(\ref{Th2d}) that $\beta_1$ and $\beta_2$ depend on $\alpha_1$ and $\alpha_2$ only linearly, which is a big advantage.

Now, we substitute equations (\ref{Th2c})-(\ref{Th2d}) into (\ref{Th2a}). After simplification, we obtain a quadratic equation for the ratio $h\equiv r_2/r_1$ as
\[ \label{Eq:h}
ah^2+bh+c=0,
\]
where the coefficients are
\begin{eqnarray*} 
&& a=1-(1+g)r_0^2, \quad b=2gpr_0\cos(\gamma_0+\gamma_1-\gamma_2),   \\
&& \hspace{2cm} c=-\left[1+(g-1)r_0^2\right].
\end{eqnarray*}
After this $h$ value is obtained, we insert (\ref{Th2c}) into the equation $p=|\alpha_1|^2+|\beta_1|^2$ and use it to obtain $r_1$ as
\[
r_1=\sqrt{\frac{p}{\Omega}},  \nonumber
\]
where
\[
\Omega=1+g^2p^2+(1+g)^2r_0^2h^2-2g(1+g)pr_0h\cos(\gamma_0+\gamma_1-\gamma_2),   \nonumber
\]
and the $r_2$ value is then
\[
r_2=r_1h.     \nonumber
\]
By now, the $\alpha_1$ and $\alpha_2$ values have been obtained, with their phases $\gamma_1,\gamma_2$ being free constants, and their amplitudes $r_1, r_2$ related to their phases and $q$ through the above equations. The $\beta_1, \beta_2$ values are determined subsequently from $\alpha_1, \alpha_2, p$ and $q$ through equations (\ref{Th2c})-(\ref{Th2d}). We have verified that the $\alpha_1, \beta_1, \alpha_2, \beta_2$ values thus obtained satisfy the condition $\alpha_1^*\alpha_2+\beta_1^*\beta_2= q$; thus the calculations are consistent.

The existence and number of solutions to equations (\ref{Th2a})-(\ref{Th2d}) depend on the existence and number of non-negative solutions to the  quadratic equation (\ref{Eq:h}) for $h$. The discriminant $\Delta=b^2-4ac$ of this quadratic equation can be found to be
\[
\Delta=4g^2p^2r_0^2\left[\cos^2(\gamma_0+\gamma_1-\gamma_2)-1-\frac{r_0^2-1}{r_0^2[1+(g^2-1)r_0^2]}\right].   \nonumber
\]
Without loss of generality, we let $0<\eta_1<\eta_2$; hence $g>1$. Then, utilizing this discriminant and the coefficient expressions of $(a,b,c)$ above, we can easily reach the following conclusions.
\begin{enumerate}
\item If $r_0>1$, then $\Delta<0$. In this case, the quadratic equation (\ref{Eq:h}) for $h$ does not admit any non-negative solution.
\item If $r_0=1$, then $p=1$, $a=c=-g$, and $b=2g\cos(\gamma_0+\gamma_1-\gamma_2)$. In this case,
the quadratic equation (\ref{Eq:h}) admits a single (repeated) positive root $h=1$ when $\cos(\gamma_0+\gamma_1-\gamma_2)=1$, and the corresponding $\textbf{w}_{10}$ and $\textbf{w}_{20}$ solutions are
\begin{eqnarray*}
&& \textbf{w}_{10}=[2^{-1/2}e^{i\gamma_1}, 2^{-1/2}e^{i\gamma_1}, 1]^T, \\  
&& \textbf{w}_{20}=[2^{-1/2}e^{i\gamma_2}, 2^{-1/2}e^{i\gamma_2}, 1]^T,
\end{eqnarray*}
where $\gamma_1$ and $\gamma_2$ are free real constants.
\item If $1/\sqrt{1+g}<r_0<1$, then this quadratic $h$-equation admits two positive solutions when
\[ 
\cos(\gamma_0+\gamma_1-\gamma_2) > \sqrt{1+\frac{r_0^2-1}{r_0^2\left[1+(g^2-1)r_0^2\right]}},     \nonumber
\]
and thus there are two $(\textbf{w}_{10}, \textbf{w}_{20})$ solutions. When the left and right sides of the above inequality become equal,
there is a single $(\textbf{w}_{10}, \textbf{w}_{20})$ solution.
\item If $r_0< 1/\sqrt{1+g}$, then $c/a<0$. In this case, the quadratic equation (\ref{Eq:h}) admits a single positive root $h$ for arbitrary $\gamma_0, \gamma_1$ and $\gamma_2$ values. Thus, there is a single $(\textbf{w}_{10}, \textbf{w}_{20})$ solution for arbitrary free parameters $\gamma_0, \gamma_1, \gamma_2$.
\end{enumerate}
To summarize, the above results reveal that equations (\ref{Th2c})-(\ref{Th2d}) admit solutions for $\textbf{w}_{10}$ and $\textbf{w}_{20}$ if and only if $|\alpha_1^*\alpha_2+\beta_1^*\beta_2|\le 1$, and the admitted solutions have four free real parameters, which can be chosen as the amplitude and phase of parameter $q=\alpha_1^*\alpha_2+\beta_1^*\beta_2$, and the phases of complex numbers $\alpha_1, \alpha_2$.

Next, we illustrate the dynamics of these two-solitons with imaginary eigenvalues. We will fix $\eta_1=i$ and $\eta_2=2i$ and vary the free parameters $q$ and phases $\gamma_1, \gamma_2$ of $\alpha_1, \alpha_2$. For these $\eta_1$ and $\eta_2$ values, $g=3$.

First, we choose
\[ \label{para1}
q=0, \quad \gamma_1=1, \quad \gamma_2=2.
\]
For this $q$ value, $r_0<1/\sqrt{1+g}$. Thus, it belongs to the case (4) above, and there is a single solution for $(\alpha_1, \beta_1, \alpha_2, \beta_2)$, which is found to be
\[ 
\alpha_1=\frac{1}{\sqrt{6}}e^{i}, \hspace{0.1cm} \alpha_2=\frac{1}{\sqrt{6}}e^{2i}, \hspace{0.1cm}
\beta_1=-\alpha_1, \quad \beta_2=\alpha_2.     \nonumber
\]
The corresponding $u(x,t)$ solution from Eq. (\ref{uvmanakov}) is displayed in Fig. 2(b). It is seen that this two-soliton meanders  periodically, which is an interesting and distinctive pattern. Physically, this meandering can be understood through the connection of the nonlocal NLS equation (\ref{RXNLS}) with the Manakov system (\ref{Manakovu})-(\ref{Manakovv}). Specifically, the evolution in Fig. 2(b) corresponds to an interaction between this $u(x,t)$ component and its opposite-parity wave $u(-x,t)$ in the $v$-component in the Manakov system. Thus, this interesting meandering of the $u(x,t)$ solution is caused by the interference of its opposite-parity wave $u(-x,t)$. Note that this meandering in Fig. 2(b) resembles internal oscillations of vector solitons in the coupled NLS equations \cite{Yangvectorsoliton}. However, in contrast with the internal oscillations reported in \cite{Yangvectorsoliton}, the present meandering does not emit any radiation and thus lasts forever. In addition, the present meandering is described by exact analytical formulae.

Next, we choose
\[ \label{para2}
q=0.6e^{3i}, \quad \gamma_1=1.5, \quad \gamma_2=5.
\]
For this $q$ value, $1/\sqrt{1+g}<r_0<1$. Thus, it belongs to case (3) above. It is easy to check that the inequality condition in case (3) is met. Hence, there are two sets of $(\alpha_1, \beta_1, \alpha_2, \beta_2)$ values.
The corresponding two $u(x,t)$ solutions from Eq. (\ref{uvmanakov}) are displayed in Fig. 2(c,d) respectively. The solution in panel (c) looks like a periodic wave drifting and recovering, while the solution in panel (d) looks like asymmetric meandering.

\begin{figure}[htbp]
\begin{center}
\includegraphics[width=0.48\textwidth]{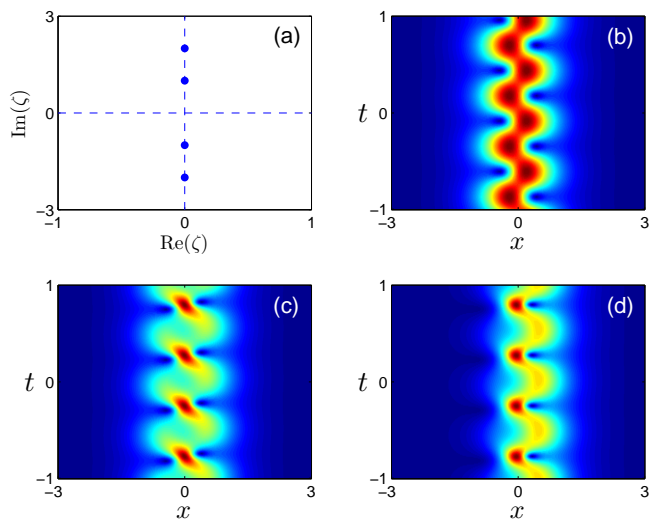}
\caption{Three examples of two-solitons in the nonlocal NLS equation (\ref{RXNLS}) with purely imaginary eigenvalues $\eta_1=i$ and $\eta_2=2i$. (a) Positions of eigenvalues; (b) the two-soliton with parameters in Eq. (\ref{para1}); (c, d) the two solutions of two-solitons with parameters in Eq. (\ref{para2}). } \label{f:fig2}
\end{center}
\end{figure}

\subsection{Two-solitons with non-imaginary eigenvalues}
\begin{figure}[htbp]
\begin{center}
\includegraphics[width=0.48\textwidth]{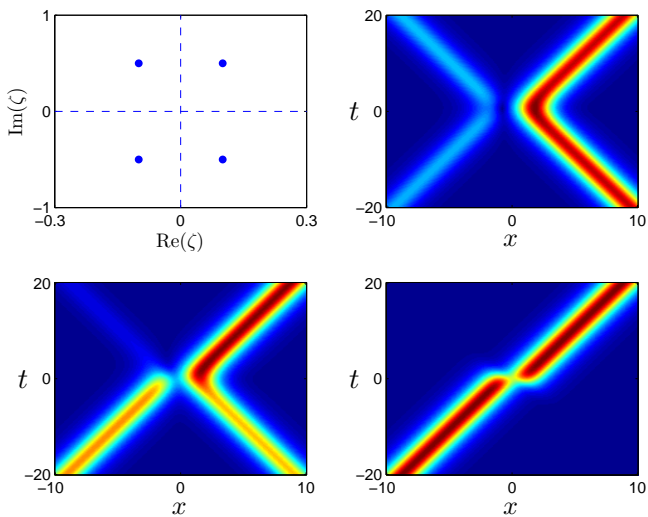}
\caption{Three examples of two-solitons in the nonlocal NLS equation (\ref{RXNLS}) with complex eigenvalues $\zeta_1=-\zeta_2^*=0.1+0.5i$ and
$\beta_1=-0.43$. Upper left: positions of eigenvalues; upper right: $\alpha_1=0.08-0.12i$; lower left: $\alpha_1=0.04$; lower right: $\alpha_1=0$. } \label{f:fig3}
\end{center}
\end{figure}

Now we consider two-solitons with non-imaginary eigenvalues, which are obtained from the two-Manakov-solitons (\ref{uvmanakov}) with a pair of non-imaginary eigenvalues $(\zeta_1, -\zeta_1^*)$ in $\mathbb{C}_+$, and with eigenvectors $\textbf{w}_{10}$, $\textbf{w}_{20}$ satisfying the equations (\ref{wsym2c}) in Theorem \ref{Theorem1}. In these solutions, $\alpha_1$ and $\beta_1$ are free complex parameters. To illustrate, we take
\[
\zeta_1=0.1+0.5i, \quad \beta_1=-0.43.   \nonumber
\]
Then, for three choices of the $\alpha_1$ values of $0.08-0.12i$, $0.04$ and $0$, the corresponding $u(x,t)$ solutions are displayed in Fig. \ref{f:fig3}. The solution in the upper right panel looks like a refection of two moving waves of different amplitudes. The solution in the lower left panel looks like the annihilation of the left-moving wave by the right-moving one upon collision. The solution in the lower right panel looks like a single right-moving wave, with its position abruptly shifted near $x=0$. Again, these interesting behaviors can be understood physically through the connection of the nonlocal NLS equation (\ref{RXNLS}) with the Manakov system (\ref{Manakovu})-(\ref{Manakovv}). For instance, the abrupt position shift of the single right-moving wave in the lower right panel is caused by a collision of this right-moving wave $u(x,t)$ with its opposite-parity wave $u(-x,t)$ in the $v$-component, which occurs near $x=0$. It is interesting to note that for the original nonlocal defocusing NLS equation proposed in \cite{AblowitzMussPRL2013}, single moving dark solitons with abrupt position shifts were reported in \cite{Xu2015}. Although such dark solitons with abrupt position shifts were derived mathematically, they were difficult to understand physically. In view of the moving bright solitons with abrupt position shifts in Fig. \ref{f:fig3}, those dark solitons with abrupt position shifts are now a little easier to understand.

Recall from Sec. \ref{onesoliton} that one-solitons in the underlying nonlocal equation (\ref{RXNLS}) are stationary. Thus, these two-solitons in Fig. \ref{f:fig3} definitely are not nonlinear superpositions of those stationary one-solitons. This behavior resembles that in the previous nonlocal NLS equation (\ref{e:NLSRX}) as we revealed in \cite{YangNsoliton}.

\section{Other new integrable nonloal equations}
Extending the idea of previous sections, we can derive other new nonlocal equations of physical relevance.

Starting from the Manakov system (\ref{Manakovu})-(\ref{Manakovv}), when we impose the solution constraint
\[  \label{constraint2}
v(x,t)=u^*(x, -t),
\]
we get
\[ \label{RTNLS}
iu_t(x,t)+u_{xx}(x,t)+2\sigma\left[|u(x,t)|^2+|u(x,-t)|^2\right]u(x,t)=0,
\]
which is a new nonlocal NLS equation of reverse-time type. When we impose the solution constraint
\[  \label{constraint3}
v(x,t)=u^*(-x, -t),
\]
the Manakov system reduces to
\[ \label{RXTNLS}
iu_t(x,t)+u_{xx}(x,t)+2\sigma\left[|u(x,t)|^2+|u(-x,-t)|^2\right]u(x,t)=0,
\]
which is a new nonlocal NLS equation of reverse-space-time type. These two equations differ from the previous nonlocal NLS equations of reverse-time and reverse-space-time types in \cite{AblowitzMussSAPM} in the nonlinear terms. Both equations are also integrable, and their Lax pairs are (\ref{Lax1})-(\ref{Lax2}) with $v(x,t)$ replaced by $u^*(x, -t)$ and $u^*(-x, -t)$ respectively.

Physically, the reverse-time NLS equation (\ref{RTNLS}) describes the solutions of the Manakov system under special initial conditions where $v(x, 0)=u^*(x, 0)$. In this case, the solution $u(x,t)$ of the reverse-time equation (\ref{RTNLS}) for negative time gives the $v(x,t)$ solution of the Manakov system for positive time through $v(x,t)=u^*(x, -t)$. The reverse-space-time NLS equation (\ref{RXTNLS}) describes the solutions of the Manakov system under special initial conditions where $v(x, 0)=u^*(-x, 0)$. In this case, the solution $u(x,t)$ of the reverse-space-time equation (\ref{RXTNLS}) for negative time gives the $v(x,t)$ solution of the Manakov system for positive time through $v(x,t)=u^*(-x, -t)$.

The above ideas can be generalized further. For instance, let we consider the four-component coupled NLS equations
\[ \label{Umodel}
iU_t+U_{xx}+2\sigma(U^\dagger U)U=0,
\]
where $U=[u, v, w, s]^T$, and $\sigma=\pm 1$. These coupled equations govern the nonlinear interaction of four incoherent light beams \cite{Kivsharbook} as well as the evolution of four-component Bose-Einstein condensates \cite{BEC}. These equations are also integrable \cite{Yang2010,Ablowitz_Trubatch_book}. If we impose the solution constraints
\[ \label{constraint4}
w(x,t)=u(-x,t), \quad s(x,t)=v(-x,t),
\]
these equations reduce to
\begin{eqnarray}
&& \hspace{-0.8cm}   iu_t(x,t)+u_{xx}(x,t)+2\sigma \left[|u(x,t)|^2+|u(-x,t)|^2 \right.    \nonumber \\
&& \hspace{1cm} \left. +|v(x,t)|^2+|v(-x,t)|^2\right]u(x,t)=0,     \\
&& \hspace{-0.8cm}  iv_t(x,t)+v_{xx}(x,t)+2\sigma \left[|u(x,t)|^2+|u(-x,t)|^2 \right.    \nonumber \\
&& \hspace{1cm} \left. +|v(x,t)|^2+|v(-x,t)|^2\right]v(x,t)=0,
\end{eqnarray}
which are a system of nonlocal Manakov equations of reverse-space type. If we impose the solution constraints
\[ \label{constraint5}
w(x,t)=u^*(x,-t), \quad s(x,t)=v^*(x,-t),
\]
we get
\begin{eqnarray}
&& \hspace{-0.8cm}  iu_t(x,t)+u_{xx}(x,t)+2\sigma \left[|u(x,t)|^2+|u(x,-t)|^2 \right.    \nonumber \\
&& \hspace{1cm} \left. +|v(x,t)|^2+|v(x,-t)|^2\right]u(x,t)=0,     \\
&& \hspace{-0.8cm}  iv_t(x,t)+v_{xx}(x,t)+2\sigma \left[|u(x,t)|^2+|u(x,-t)|^2 \right.    \nonumber \\
&& \hspace{1cm} \left. +|v(x,t)|^2+|v(x,-t)|^2\right]v(x,t)=0,
\end{eqnarray}
which are a system of nonlocal Manakov equations of reverse-time type. If we impose the solution constraints
\[ \label{constraint6}
w(x,t)=u^*(-x,-t), \quad s(x,t)=v^*(-x,-t),
\]
we get
\begin{eqnarray}
&& \hspace{-0.8cm} iu_t(x,t)+u_{xx}(x,t)+2\sigma \left[|u(x,t)|^2+|u(-x,-t)|^2 \right.    \nonumber \\
&& \hspace{1cm} \left. +|v(x,t)|^2+|v(-x,-t)|^2\right]u(x,t)=0,    \\
&& \hspace{-0.8cm}  iv_t(x,t)+v_{xx}(x,t)+2\sigma \left[|u(x,t)|^2+|u(-x,-t)|^2 \right.    \nonumber \\
&& \hspace{1cm} \left. +|v(x,t)|^2+|v(-x,-t)|^2\right]v(x,t)=0,
\end{eqnarray}
which are a system of nonlocal Manakov equations of reverse-space-time type. These three nonlocal Manakov systems are also integrable, and they describe the solution behaviors of the physical model (\ref{Umodel}) under special initial conditions of (\ref{constraint4}), (\ref{constraint5}) and (\ref{constraint6}) with $t=0$.

\section{Summary and discussion}
In this paper, we proposed a new integrable nonlocal NLS equation (\ref{RXNLS0}) which has concrete physical meanings. This equation was derived from a reduction of the Manakov system, and it describes physical situations governed by the Manakov system under special initial conditions. Solitons and multi-solitons in this nonlocal equation were also investigated in the framework of Riemann-Hilbert formulation. We found that symmetry relations of discrete scattering data for this nonlocal equation are very complicated, which makes the derivation of its general $N$-solitons challenging. From the one- and two-solitons we obtained. it was observed that the two-solitons are not a nonlinear superposition of one-solitons, and the two-solitons exhibit interesting dynamical patterns such as meandering and abrupt position shifts. As a generalization of these results, we also proposed other integrable and physically meaningful nonlocal equations, such as new NLS equations of reverse-time and reverse-space-time types, as well as nonlocal Manakov equations of reverse-space, reverse-time and reverse-space-time types.

The results in this paper are significant in two different ways. From a mathematical point of view, we presented a new integrable nonlocal equation which has clear physical meanings. In addition, we showed that this integrable equation exhibits some unusual mathematical properties such as intricate symmetry relations of its discrete scattering data. From a physical point of view, we derived one- and two-solitons in this nonlocal equation, which correspond to Manakov solutions under the initial parity symmetry between the two components, and these solitons feature interesting physical patterns such as symmetric and asymmetric meandering.

The highly complex symmetry relations of discrete scattering data for the nonlocal NLS equation (\ref{RXNLS0}) are very surprising. This fact implies that general $N$-solitons in this equation will be very difficult to derive in the inverse scattering and Riemann-Hilbert framework. Whether they can be derived more easily in other frameworks such as the Darboux transformation and bilinear methods remains to be seen.

In this paper, we only studied bright solitons in the nonlocal NLS equation (\ref{RXNLS0}). Other types of solutions such as rogue waves and dark solitons in this equation are desirable too, which merit studies in the future. In addition, we proposed a number of other new nonlocal equations of physical relevance, such as new NLS equations of reverse-time and reverse-space-time types, and nonlocal Manakov equations of reverse-space, reverse-time and reverse-space-time types. Bright solitons, dark solitons and rogue waves in those systems are also open questions for further studies.

\section*{Acknowledgment}
This material is based upon work supported by the Air Force Office of Scientific Research under award number FA9550-18-1-0098, and the National Science Foundation under award number DMS-1616122.

\end{document}